\documentclass[preprint,pra,nofootinbib]{revtex4}
\usepackage[mathscr]{euscript}
\usepackage{graphicx}
\usepackage{float,epsfig}
\usepackage{dcolumn}
\usepackage{bm}
\usepackage{graphicx}
\usepackage{amsmath,amssymb,amsthm}
\usepackage{subfigure}
\usepackage{color}
\usepackage[colorlinks=true,linkcolor=blue]{hyperref}
\usepackage[normalem]{ulem}
\usepackage[makeroom]{cancel}

\newcommand{\bea}{\begin{eqnarray}}
\newcommand{\eea}{\end{eqnarray}}
\newcommand{\beq}{\begin{equation}}
\newcommand{\eeq}{\end{equation}}

\def\/{\over}

\newcommand{\rev}[1]{\textcolor{red}{{#1}}}
\newcommand{\vk}[1]{\textcolor{blue}{{#1}}}

\begin{document}

\title{Localization and delocalization of light in photonic moir\'e lattices}

\author{Peng Wang,${}^{1}$ Yuanlin Zheng,${}^{1}$ Xianfeng Chen,${}^{1}$ Changming Huang,${}^{2}$ Yaroslav V. Kartashov,${}^{3,4}$ Lluis Torner,${}^{3,5}$ Vladimir V. Konotop,${}^{6}$ Fangwei Ye$^{1\ast}$\\
\emph{${}^{1}$ School of Physics and Astronomy, Shanghai Jiao Tong University, Shanghai 200240, China}\\
\emph{${}^{2}$Department of Electronic Information and Physics, Changzhi University, Shanxi 046011, China}\\ 	
\emph{${}^{3}$ICFO-Institut de Ciencies Fotoniques, The Barcelona Institute of Science and Technology,}\\ 
\emph{ 08860 Castelldefels (Barcelona), Spain}\\
\emph{${}^{4}$Institute of Spectroscopy, Russian Academy of Sciences, Troitsk, Moscow, 108840, Russia}\\
\emph{${}^{5}$Universitat Politecnica de Catalunya, 08034 Barcelona, Spain}\\
\emph{${}^{6}$Departamento de F\'{i}sica, and Centro de F\'{i}sica Te\'{o}rica e Computacional,}\\
\emph{Faculdade de Ci\^encias, Universidade de Lisboa, Campo Grande, Ed. C8, Lisboa 1749-016, Portugal} 
\\
\emph{$^\ast$Corresponding author:  fangweiye@sjtu.edu.cn}
}

\date{\today}

%

\maketitle

{\bf { \bf Moir\'e lattices consist of two identical periodic structures overlaid with a relative rotation angle. Observed even in everyday life, moir\'e lattices have been also produced with coupled graphene-hexagonal boron nitride monolayers~\cite{GrapheneMoire,Woods14,Thermal}, graphene-graphene layers~\cite{BandFlatGraphene,UnconvenSupercond}, and layers on a silicon carbide surface~\cite{Science}. The recent surge of interest in moir\'e lattices is connected with a possibility to explore in such structures a rich variety of unusual physical phenomena, such as commensurable-incommensurable transitions and topological defects~\cite{Woods14}, emergence of insulating states due to band flattening~\cite{BandFlatGraphene,flat_band}, unconventional superconductivity~\cite{UnconvenSupercond} controlled by the rotation angle between monolayers~\cite{magic,flat_hexagon2}, quantum Hall effect~\cite{Dean13}, realization of non-Abelian gauge potentials~\cite{Gauge}, as well as quasicrystals appearing at special rotation angles~\cite{Stampfli}. However, a fundamental question that remains unexplored is the  evolution of waves in the potentials defined by the moir\'e lattices. Here we {experimentally create} two-dimensional photonic moir\'e lattices, which, unlike their crystalline predecessors, have controllable parameters and symmetry {allowing to explore} transitions between structures with fundamentally different periodicity: periodic, general aperiodic and quasi-crystal ones. Equipped with such realization, we observe localization of light in deterministic linear lattices. \rev{Such localization is based on the flat-band physics~\cite{flat_band}, thus contrasting} with previous schemes based on light diffusion in optical quasicrystals~\cite{Freeman}, where disorder is required~\cite{Levi} for the onset of Anderson localization~\cite{Anderson2D}. Using commensurable and incommensurable moir\'e patterns, we also provide the first experimental demonstration of two-dimensional localization-delocalization-transition (LDT) of wavepacket in an optical setting. Moir\'e lattices may feature almost arbitrary geometry consistent with the crystallographic symmetry groups of the sublattices, they allow comprehensive exploration of the physics of transitions between periodic and aperiodic phases and offer an alternative way for manipulating light by light.}}


One of the most important properties of an engineered optical system is its capability to affect the light beam in a prescribed manner, in particular to localize it or to control its diffraction. The importance of the problem of wave localization extends far beyond optics. It was studied in all branches of physics dealing with wave phenomena. In particular, it is well known that homogeneous or strictly periodic linear systems cannot result in localization of waves, and that inhomogeneities, random or regular, or nonlinear effects are required for localization. Wave localization in a random medium, alias Anderson localization~\cite{Anderson}, is a hallmark discovery of the condensed-matter physics. All electronic states in one- and two-dimensional potentials with uncorrelated disorder are localized. Three-dimensional systems with disordered potentials are known to have both localized and delocalized eigenstates~\cite{Anderson2D}, separated by an energy known as the mobility edge~\cite{mobility}.

\rev{If disorder is correlated, a mobility edge may exist in one-dimensional systems too, as it was observed experimentally for a Bose-Einstein condensate in a speckle potential~\cite{Aspect}. Furthermore, coexistence} of localized and delocalized eigenstates has been predicted also in regular quasiperiodic one-dimensional systems, first in the discrete Aubri-Andr\'e~\cite{AA} model and later in continuous optical and matter-wave systems~\cite{Boers,Modugnho,LocDelocSoc}. Quasiperiodic (or aperiodic) structures, even those that possess long-range order, fundamentally differ both from periodic systems where all eigenmodes are delocalized Bloch waves, and from disordered media where all states are localized (in one or two dimensions). Upon variation of the  parameters of a quasiperiodic system, it is possible to observe transition between localized and delocalized states. Such LDT has been observed in one-dimensional quasiperiodic optical~\cite{Lahini} and in atomic systems~\cite{AL_BEC_bichromat,Luschen18}.

Wave localization is sensitive to dimensionality of the problem, irrespective of the type of the medium. \rev{Anderson localization and mobility edge in two-dimensional disordered systems were first reported in the experiment with bending waves~\cite{Sheng} and later in optically induced disordered lattices~\cite{anderson1}. In quasicrystals localization has been observed} only under the action of nonlinearity~\cite{Freeman} {and} in the presence of strong disorder~\cite{Levi}. Localization and delocalization of light in two-dimensional systems without any type of disorder and nonlinearity, have never been observed so far.

Here we report on the first experimental realization of reconfigurable photonic moir\'e lattices with controllable parameters and symmetry. The lattices are induced by two superimposed periodic patterns~\cite{Fleischer,GauM} (sublattices) with either square or hexagonal primitive cells~\cite{LDTSR}. The sublattices of each moir\'e pattern have tunable amplitudes and twist angle. Depending on the twist angle a photonic moir\'e lattice may have different periodic (commensurable) structure or aperiodic (incommensurable) structure without translational periodicity, but they always feature the same rotational symmetry as the symmetry of the sublattices. Moir\'e lattice can also transform into quasicrystals with higher rotational symmetry (e.g. the case of the Stampfli pattern~\cite{Stampfli}). 
The angles at which a commensurable phase (periodicity) of an optically-induced moir\'e lattice is achieved, are determined by the Pythagorean triples in the case of square sublattices or by another Diophantine equation, when the primitive cell of the sublattices is not square (see Methods). For all other rotation angles the structure is aperiodic albeit regular (i.e., it is not disordered). Importantly, changing the relative amplitudes of the sublattices allows to smoothly tune the shape of the lattice without affecting its rotational symmetry. In contrast to crystalline moir\'e lattices~\cite{GrapheneMoire,Woods14,Thermal,BandFlatGraphene,UnconvenSupercond, Science}, optical patterns are monolayer structures, i.e., both sublattices interfere in one plane. As a consequence, light propagating in such media is described by a one-component field [see Eq.~(\ref{equation1}) below], rather than by the spinors characterizing electronic states in the tight-binding approximation applicable for material moir\'e lattices~\cite{magic,flat_hexagon2}.

In the paraxial approximation, the propagation of an extraordinary polarized light beam in a photorefractive medium with optically induced refractive index landscape is governed by the Schr\"{o}dinger equation for the dimensionless signal field amplitude $\psi({\bf r},z)$ \rev{\cite{Fleischer,Efremidis}}:
\begin{equation}
	\label{equation1}
	\textit{i}\frac{\partial {\psi}}{\partial \textit{z}}=-\frac{1}{2}\nabla^2\psi+\frac{E_0}{1+I({\bf r})} {\psi}.   
\end{equation}
Here $\nabla=(\partial/\partial x,\partial/\partial y) $; ${\bf r}=(x,y)$ is the radius-vector in the transverse plane, which  is scaled to the wavelength $\lambda=632.8~\textrm{nm}$ of the beam used in our experiments; $z$ is the propagation distance scaled to the characteristic diffraction length $2\pi n_{\rm e} \lambda$;  $n_{\rm e}$ is the refractive index of the homogeneous crystal for extraordinary-polarized light, $E_0>0$ is the dimensionless applied dc field; $I({\bf r})\equiv \left|p_1V({\bf r})+p_2V(S{\bf r})\right|^2$ is the moir\'e lattice induced by two ordinary polarized {mutually coherent} periodic sublattices $V({\bf r})$ and $V(S{\bf r})$ interfering in the $(x,y)$ plane and rotated by the angle $\theta$ with respect to each other (see Methods and~\cite{Fleischer} for details of the optical induction technique); $S=S(\theta)$ is the operator of the two-dimensional rotation; $p_1$ and $ p_2$ are the amplitudes of the first and second sublattices, respectively. The number of laser beams creating each sublattice $V({\bf r})$ depends on the desired lattice geometry. The form in which the lattice intensity $I({\bf r})$ enters Eq.~(\ref{equation1}) is determined by the mechanism of photorefractive response.

To visualize the formation of moir\'e lattices it is convenient to associate a continuous sublattice $V({\bf r})$ with a discrete one with lattice vectors determined by the locations of the absolute maxima of $V({\bf r})$. The resulting moir\'e pattern inherits the rotational symmetry of $V({\bf r})$. At specific angles some nodes of different sublattices may coincide thereby leading to translational symmetry of the moir\'e patterns $I({\bf r})\equiv \left|p_1V({\bf r})+p_2V(S{\bf r})\right|^2$. Here we focus on square and hexagonal sublattices $V({\bf r})$, whose primitive translation vectors, in commensurable phases, are illustrated by blue arrows, respectively, in Fig. 1 (the first and third columns) and in Fig. 4 (the first and second columns). The rotation angles at which the periodicity of $I({\bf r})$ is achieved are determined by triples of positive integers $(a,b,c)\in\mathbb{Z}^+$ related by a Diophantine equation characteristic for a given sublattice~\cite{LDTSR} (see Methods). 
 
\begin{figure}[htb]
\centering
\includegraphics[width=0.8\textwidth]{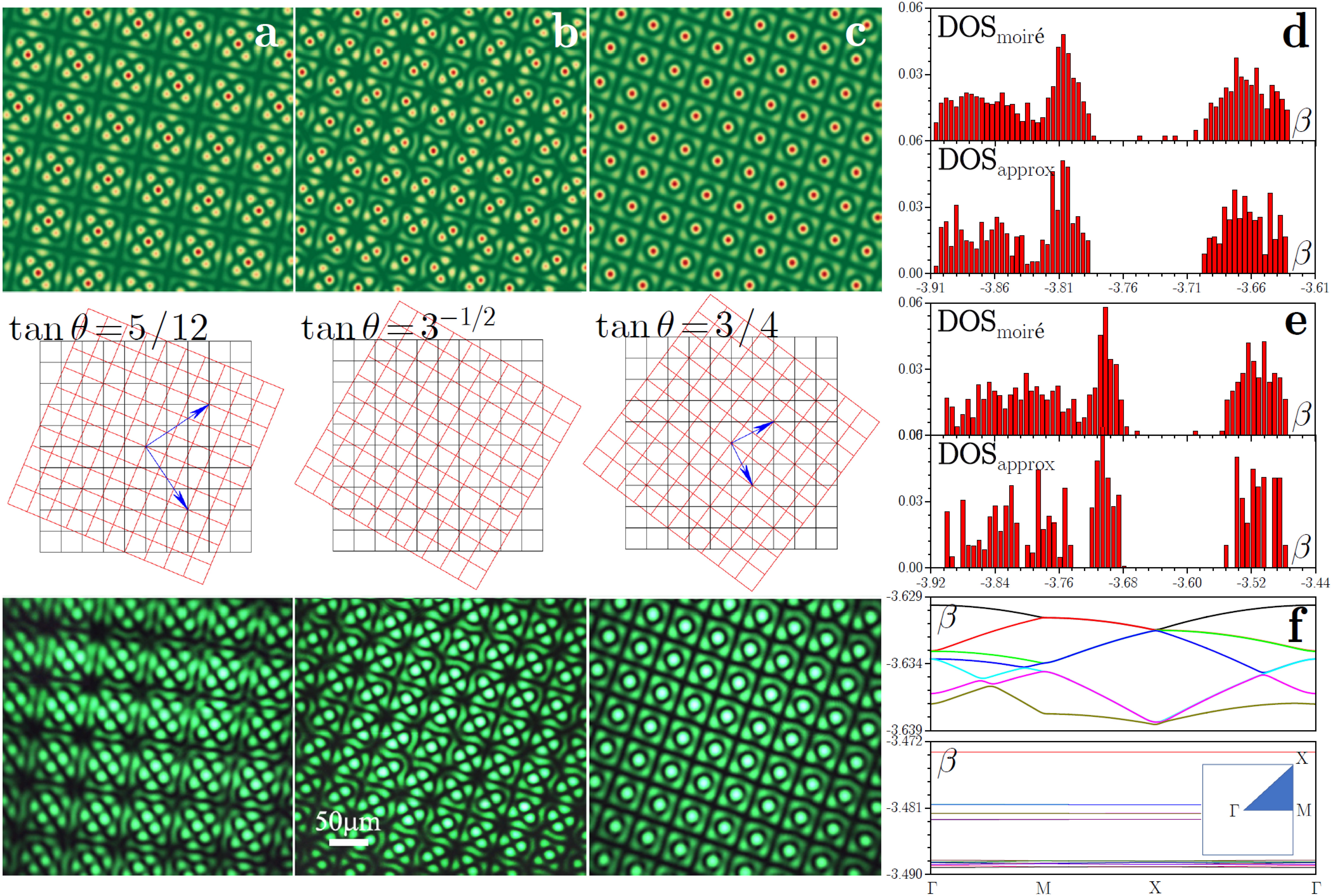}
\caption{(a)-(c) Moir\'e lattices $ \textit{I}\left({\bm r} \right) $ generated by two interfering square sublattices with $ {p}_{1}={p}_{2} $, whose axes are mutually rotated by the angle indicated in each panel. First row: calculated patterns. Second row: schematic discrete representation of two rotated sublattices. Third row: experimental patterns observed at the output face of the crystal. The scale is the same for all experimental images. \rev{Comparison of DOS calculated for moir\'e lattice (top) and its periodic approximation (bottom) at $p_2=0.1$ (d) and $p_2=0.2$ The approximate Pythagorean lattice has the lattice vectors} \vk{{\bf $b_1=...$}} \rev{ (see Supplementary Information), (e). (f) Band structures for periodic lattice approximating moir\'e lattice at $p_2=0.1$ (top, 7 upper bands are shown) and $p_2=0.2$ (bottom, 18 upper bands are shown). In all cases $p_1=1$.}
}
\label{fig:one}
\end{figure}

First, we consider a Pythagorean lattice created by two square sublattices. For the rotation angles $\theta$, such that $\cos\theta=a/c$ and $\sin\theta=b/c$, where $(a,b,c)$ is a Pythagorean triple, i.e., $a^2+b^2=c^2$, $I({\bm r})$ a is fully periodic (commensurable) moir\'e lattice. {Such angle is referred below as Pythagorean}. For all other rotation {non-Pythagorean} angles $\theta$, the lattice $I({\bm r})$ is aperiodic (incommensurable). {Figures 1(a)-(c) compare} calculated $I({\bm r})$ patterns with $p_1=p_2$ (first row) with lattices created experimentally in a biased SBN:61 photorefractive crystal with dimensions $5\times5\times20$ \text{mm}$^3$ (third row) for different rotation angles indicated on the panels. \rev{The lattice was created using optical induction technique, invented in~\cite{Efremidis} and first realized experimentally in \cite{Fleischer}.} The second row shows the respective discrete moir\'e lattices. {Columns (a) and (c)} show periodic lattices, while {column (b)} gives an example of an aperiodic lattice. All results here and below were obtained for $E_0=7$, which corresponds to a $8\times 10^4~\text{V/m}$ dc electric field applied to the crystal. The amplitude of the first sublattice was set to $p_1=1$ in all cases, which  corresponds to an average intensity $I_{\rm av}\approx  3.8~\text{mW}/\text{cm}^2$. For such parameters, the actual refractive index modulation depth in the moir\'e lattice is of the order of $\delta n \sim 10 ^{-4}$.

An important feature of optical moir\'e lattices that distinguishes them from their crystalline counterparts, is that rotated sublattices have tunable amplitudes. The ratio $p_2/p_1$ can be changed without affecting the symmetry of the entire pattern, except for the case $p_1=p_2$, when rotation symmetry changes, as discussed below. Thus, LDT in optical setting can be studied either by varying the twist angle $\theta$ at fixed amplitudes $p_{1,2}$, or by varying one or both amplitudes of the sublattices, at a fixed angle $\theta$.  

\rev{From the mathematical point of view, incommensurable lattices are almost periodic functions~\cite{Franklin}. Like any irrational number can be approached by a rational one~\cite{Hurwitz}, any non-Pythagorean twist angle can be approached by a Pythagorean one with any prescribed accuracy (as shown in Supplemental Information). Thus, any finite area of an incommensurable moir\'e lattice can be approached by a primitive effective cell of some periodic Pythagorean lattice, and more accurate approximation requires larger primitive cell of the Pythagorean lattice. This property (derived in Supplemental Information) is illustrated by quantitative similarities between densities of states (DOSs) calculated for an incommensurable lattice and its Pythagorean approximation and compared in Fig.~1(d,e) for different $p_2/p_1$ ratios. A remarkable property of Pythagorean lattices is extreme flattening of the higher bands that occurs when $p_2/p_1$ ratio exceeds certain threshold, as illustrated in Fig. 1(f). The number of flat bands rapidly grows with the size of the area of the primitive cell of the approximating Pythagorean lattice. Thus, an incommensurable moir\'e lattice can be viewed as the large-area limit of periodic Pythagorean lattices with extremely flat higher bands. Since flat bands support quasi-nondiffracting localized modes, the initially localized beam launched into such moir\'e lattice will remain localized. This flat-band physics of moir\'e lattices, earlier discussed for twisted bilayer graphene \cite{magic,flat_hexagon1,flat_hexagon2}, allows us to predict beam localization above some threshold $p_2/p_1$ value, whose physics is different from that of Anderson localization in random media. Furthermore, flat bands support states, which are exponentially localized  in the primitive cell. Such states can be well approximated by exponentially localized two-dimensional Wannier functions~\cite{Wannier} (see Fig. 2(c) and Supplementary Information). }

To elucidate the impact of the sublattice amplitudes and rotation angle $\theta$ on the light beam localization, we calculated the linear eigenmodes $\psi({\bm r},z)=w({\bm r})e^{i\beta z}$, where $\beta$ is the propagation constant and $w({\bm r})$ is the mode profile, supported by the moir\'e lattices. To characterize the localization of the eigenmodes we use the integral form-factor $\chi$ defined by $ \chi^2=U^{-2}\iint |\psi|^{4}d^2{\bm r}$, where $ U=\iint|\psi|^{2}d^2{\bm r}$ is the energy flow (the integration is over the transverse area of the crystal). Form-factor is inversely proportional to the mode width: the larger $\chi$ the stronger the localization. The dependence of the form-factor of the most localized mode of the structure (mode with largest $\beta$) on $\theta$ and $p_2$ is shown in Fig. 2(a) {(for modes with lower $\beta$ values these dependencies are qualitatively similar)}. One observes a sharp LDT above a certain threshold depth $p_2^{\rm LDT}$ of the second sublattice, at fixed amplitude of the first sublattice: $p_1=1$. \rev{This corroborates with strong band flattening of approximating Pythagorean lattice at $p_2>p_2^{\rm LDT}$ depicted in Fig. 1(f).} Below $p_2^{\rm LDT}$ all modes are found to be extended [Fig. 2(b)] and above the threshold, some modes are localized [Fig. 2(c)]. \rev{Inset in Fig. 2(c) reveals exponential tails for $p_2>p_2^{\rm LDT}$ from which localization length for most localized mode can be extracted.} In Fig. 2(a) one also observes sharp delocalization for angles $\theta$ determined by the Pythagorean triples, when all modes are extended regardless of $p_2$ value.

 We observe in Fig. 2(a) that $p_2^{\rm LDT}$ is practically independent of the rotation angle. This is explained by the fact that a large fraction of the power in a localized mode resides in the vicinity of a lattice maximum (i.e., at ${\bm r}={\bf 0}$). In an incommensurable phase $I({\bm r})< I({\bm 0})$ for all ${\bm r}\neq {\bm 0}$ and the optical potential locally can be approximated by the Taylor expansion of $E_0/[1+I({\bm r})]$ with respect to ${\bm r}$ near the origin. Such expansion includes the rotation angle $\theta$ only in the fourth order~\rev{(see Supplementary Information)}. Thus, locally, the optical potential can be viewed as almost isotropic.

\begin{figure}[htb]
\centering
\includegraphics[width=0.8\textwidth]{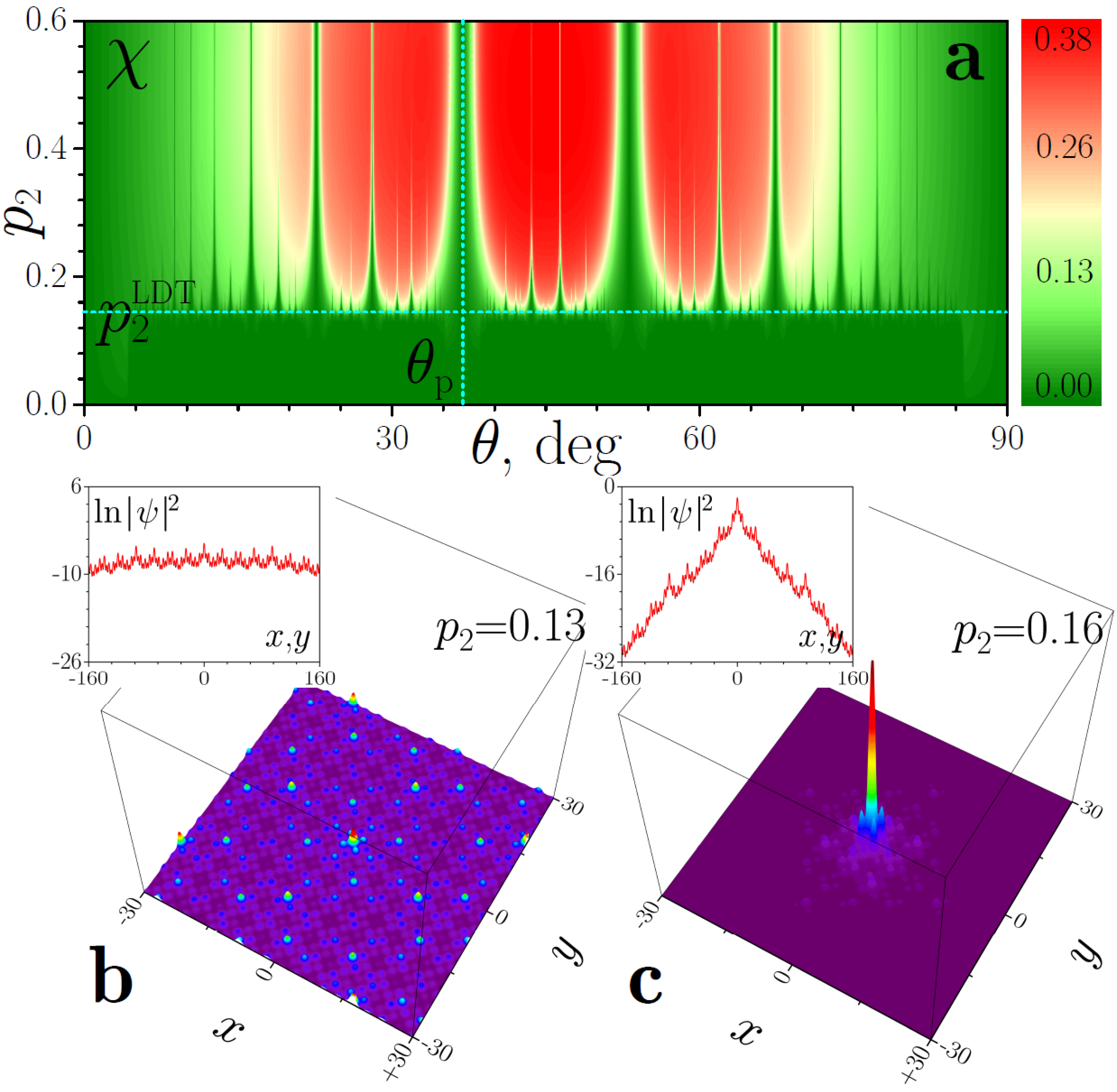}
\caption{(a) Form-factor (inverse width) of the eigenmodes with largest $\beta$ versus rotation angle $\theta $ and versus amplitude of the second sublattice $p_{2} $ at $ p_{1}=1 $. The horizontal dashed line indicates the sublattice depth $ {p}_{2}^\textrm{LDT} $ at which LDT occurs. The vertical dashed line shows one of the Pythagorean angles $ \theta_\textrm{p}=\text{arctan(3/4)} $. Examples of mode profiles with largest $\beta$ for $p_2<{p}_{2}^\textrm{LDT}$ (b) and $p_2>{p}_{2}^\textrm{LDT}$ (c). \rev{Insets show cuts of $\textrm{ln}|\psi|^2$ distribution along the $x$ and $y$ axes.}}
\label{fig:three}
\end{figure}

To study the guiding properties of the Pythagorean moir\'e lattices experimentally and to expose the two-dimensional LDT, we measured the diffraction outputs for beams propagating in lattices corresponding to different rotation angles $\theta$ \rev{for fixed input position of the beam, centered or off-center}. The diameter of the Gaussian beam focused on the input face of the crystal was about $23~\mu \textrm{m}$. Such a beam covers approximately one bright spot (channel) of the lattice profile. The intensity of the input beam was about $10$ times lower than the intensity of the lattice-creating beam, $I_{\rm av}$, to guarantee that the beam does not distort the induced refractive index and that it propagates in the linear regime. \rev{When a beam enters moir\'e lattice it either diffracts, if there are no localized modes, or it remains localized if such modes exist (notice that no averaging is used, because lattice is deterministic). Extended Data Fig.~\ref{fig:propexperiment} compares experimental and theoretical results for propagation dynamics in three different regimes at fixed $p_1=1$. In the incommensurable lattice at small $p_2<p_2^{\rm LDT}$ one observes beam broadening and no traces of localization (top row). Localization takes place at sufficiently high amplitude $p_2>p_2^{\rm LDT}$ of the second sublattice in the incommensurable case, e.g., for $p_2=1$ (middle row). However, at a twist angle $\theta$ corresponding to a commensurable moir\'e lattice, localization does not occur even for $p_2=p_1=1$ due to the restoration of lattice periodicity (bottom row). Simulations of propagation up to much larger distances beyond the available sample length presented in Extended Data Fig.~\ref{fig:propsimulation} fully confirm localization of the beam in incommensurable lattice at any distance at $p_2>p_2^{\rm LDT}$ and its expansion at $p_2<p_2^{\rm LDT}$.}


\begin{figure}[ht]
\centering
\includegraphics[width=0.5\textwidth]{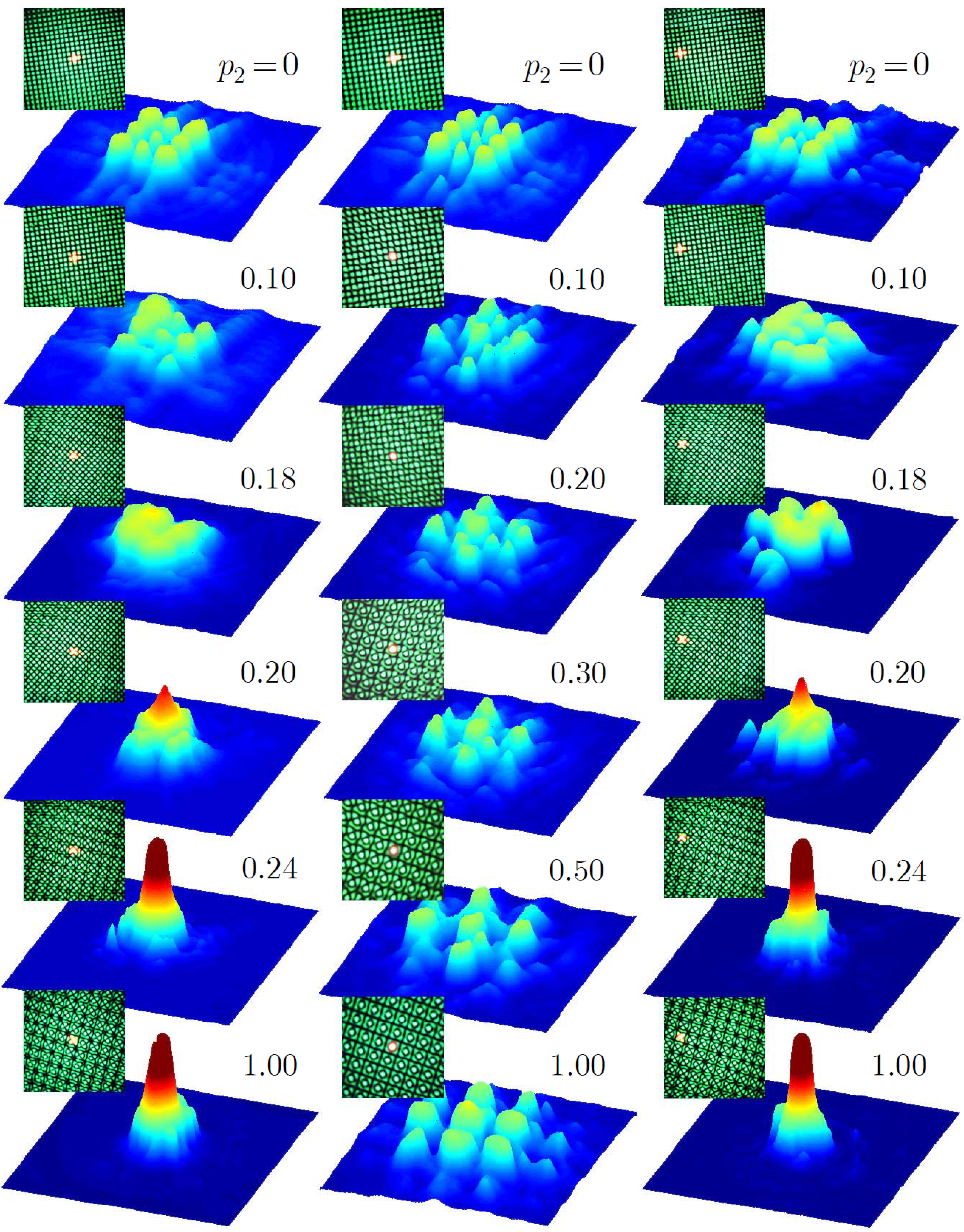}
\caption{Observed output intensity distributions illustrating LDT with increasing amplitude $p_2$ of the second sublattice for rotation angle $\theta=\arctan 3^{-1/2}=\pi/6$ (left and {right columns}) and absence of LDT for the Pythagorean angle $\theta=\arctan(3/4)$ (central column). \rev{Insets show initial excitation position, central for the left and central columns, and off-center for the right column.}}
\label{fig:four}
\end{figure}

{Extended} experimental evidence of LDT in the two-dimensional lattice is presented in Fig. 3, where we compare {output patterns} for the low-power light beam in the incommensurable ($\tan\theta=3^{-1/2}$, \rev{left and right columns for central and off-center excitations, respectively}) and commensurable ($\tan\theta=3/4$, central column) moir\'e lattices, tuning in parallel the amplitude $p_2$ of the second sublattice. When $p_2<{p}_{2}^\textrm{LDT}$  (in Fig. 3 ${p}_{2}^\textrm{LDT}\approx 0.15$), the light beam in the incommensurable lattice notably diffracts upon propagation and {expands across multiple local maxima of $I({\bm r})$ in the vicinity of the excitation point}. However, when $p_2$ exceeds the LDT threshold, it is readily visible that \rev{diffraction is arrested for both central (left column) and off-center (right column) excitations} and a localized spot is observed at the output over a large range of $p_2$ values. In clear contrast, localization is absent for any $p_2$ value in the periodic lattice associated with the Pythagorean triple (central column).

\begin{figure}[ht]
\centering
\includegraphics[width=0.9\textwidth]{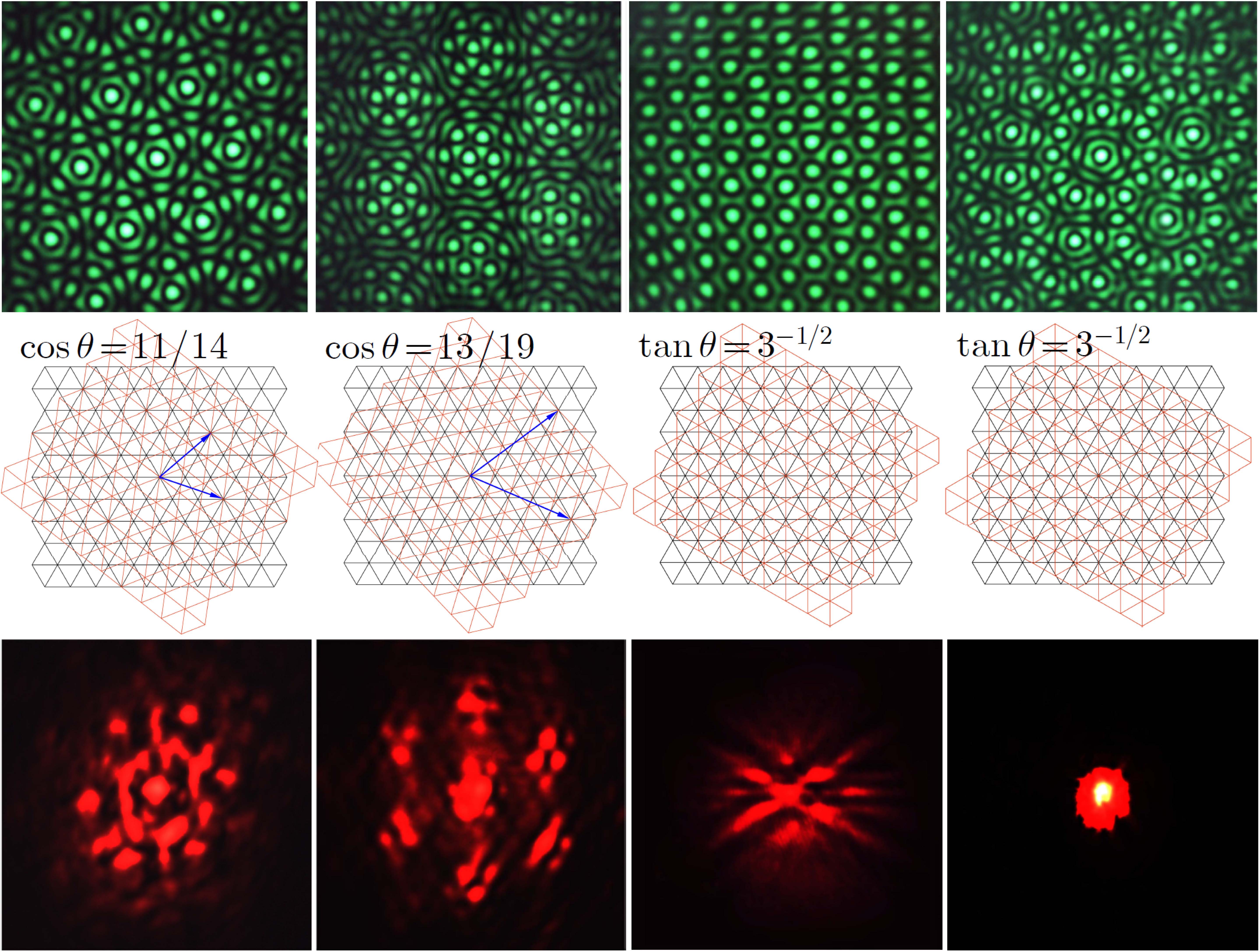}
\caption{First row: moir\'e lattices produced by interference of two hexagonal patterns rotated by the angle $\theta$: $p_2=1$ in the first, second, and fourth columns, while in the third column $p_2=0.18$. Second row: schematic discrete representation of two rotated hexagonal sublattices. Third row: measured output intensity distributions for signal beam at the output face of the crystal. In all cases $p_1=1$.}
\label{fig:five}
\end{figure}

The mutual rotation of two identical sublattices allows generation of commensurable and incommensurable moir\'e patterns with sublattices of any allowed symmetry. To illustrate the universality of LDT, we induced hexagonal moir\'e lattices \rev{(the technique of induction is similar to that used for single hexagonal photonic lattices ~\cite{photon-graphen})}. For such lattices, the rotation angles producing commensurable patterns are given by the relation $\tan \theta= b \sqrt{3}/(2a+b)$, where the integers $a$ and $b$ solve the Diophantine equation $a^2+b^2+ab=c^2$. Two examples are presented in the first and second columns of Fig. 4. In such periodic structures, the signal light beam experiences considerable diffraction for any amplitude of the sublattices, shown in the bottom row. {To} observe LDT one has to induce aperiodic structures. To such end, we considered the rotation angle of $30^o$. In such incommensurable case we did observe LDT by increasing the amplitude of the second sublattice, keeping the amplitude $p_1$ fixed. Delocalized and localized output beams are shown in the lower panels of the third and fourth columns of Fig. 4. In the third column the ideal 6-fold rotation symmetry of the output pattern is slightly distorted, presumably due to the intrinsic anisotropy of the photorefractive response. At $p_1=p_2$  the moir\'e pattern acquires the 12-fold rotational symmetry (shown in the fourth column of Fig. 4) as it was proposed in~\cite{Stampfli} as a model of a quasicrystal and is similar to the twisted bilayer graphene quasicrystal reported in~\cite{Science}.


In closing, we have created first photonic moir\'e lattices by imprinting two mutually rotated square and hexagonal sublattices into the photorefractive crystal. These structures with highly controllable parameters and symmetry have enabled systematic investigation of evolution and localization of wavepackets. \rev{Namely, we report the first experimental demonstration of two-dimensional LDT of a wavepacket in an optical setting}. Our findings uncover novel mechanism for wave localization in regular incommensurable systems \rev{based on the physics of flat-band structures}. LDT can occur according to different scenarios: via commensurable-incommensurable transition and via tuning of relative depths of optically-induced lattices. \rev{In addition to new possibilities for control of light by light, i.e. control of propagation paths, symmetry of diffraction patterns, diffraction rate of light beams, and formation of self-sustained excitations with new symmetries}, photonic moir\'e patterns allows to study phenomena relevant to other areas of physics, particularly from condensed matter systems, which are much harder to explore directly. \rev{Such lattices can be used to study   the relations between conductivity/transport and symmetry of incommensurable patterns with long-range order.} Being tunable, optical moir\'e patterns can be created in practically any arbitrary configurations consistent with two-dimensional symmetry groups, thus allowing the exploration of potentials that may not be easily produced in tunable form using material moir\'e structures. \rev{The setting reported here is also reproducible for atomic potentials, also created by the standing beams using similar geometries, and hence the effects observed here can be obtained in atomic systems, including BECs, both linear and nonlinear ones. They can be also induced in absorbing resonant atomic media, such as hot vapors. Furthermore, while most of previous studies of moir\'e lattices were focused on graphene structures and quasicrystals, nowadays widely used in material science, our results suggest that a somewhat simpler twisted square lattice is a promising alternative geometry for creation of synthetic materials allowing investigation of two-dimensional physics, including physics of localization and physics of flat-band materials.}

\paragraph*{\bf Acknowledgments} P. W and F. Y acknowledge support from the NSFC (No. 61475101 and No. 11690033). Y.V.K. and L.T. acknowledge support from the Severo Ochoa Excellence Programme (SEV-2015-0522), Fundacio Privada Cellex, Fundacio Privada Mir-Puig, and CERCA/Generalitat de Catalunya.

\paragraph*{\bf Author contributions} P. W and Y. Z contributed equally to this work. All authors contributed significantly to this work.

\paragraph*{\bf Competing interests} The authors declare no competing interests.

\section*{METHODS}
 \paragraph*{\bf Experimental setup.} The experimental setup is illustrated in Extended Data Fig. 1. A cw frequency-doubled Nd:YAG laser at wavelength $ \lambda=532~\text{nm} $ is divided by a polarizing beam splitter into two \textit{polarization} components, which are sent to Path \textbf{a} and Path \textbf{b} separately. Light in Path \textbf{a} is extraordinarily polarized and it is used to image the induced potential in the photorefractive crystal (see the third row of Fig. 1 in the main text). Light in Path \textbf{b} is ordinarily polarized and it is used to write the desirable potential landscape in the photorefractive SBN:61 crystal with dimensions $5\times5\times20~\text{mm}^{3} $ and extraordinary refractive index $n_{\rm e}=2.2817$. Before entering the crystal ordinarily polarized light beam in Path \textbf{b} is modulated by Masks 1 and 2 transforming this beam into superposition of two rotated periodic patterns. Their relative strength $p_2/p_1$, or, more precisely, the strength of the second lattice, as well as the twist angle $\theta$ are controlled by the polarizer-based Mask 1 and amplitude Mask 2.  The He-Ne laser with wavelength $\lambda=633~\text{nm}$ shown in path \textbf{c} provides extraordinarily polarized beam focused onto the front facet of the crystal, that serves as a probe beam for studying light propagation in the induced potential. We record the output light intensity pattern by a CCD at the exit facet of the crystal after propagation distance of $20~\text{mm}$.
 
{\bf Characteristics of moir\'e lattices used in experiment.} In the experiments there have been used two types of moir\'e lattices, whose characteristics are summarized  in Extended Data Table~\ref{tab:one}. In all the cases the center of  rotation in the $(x,y)$ plane was chosen coincident with a node of one of the sublattices.

\section*{DATA AVAILABILITY}

The data that support the findings of this study are available from the corresponding author upon reasonable request.

\section*{EXTENDED DATA}

 \setcounter{figure}{0}
 \renewcommand\figurename{Extended Data Fig.}
 \begin{figure}[ht]
 \centering
 \includegraphics[width=0.9\textwidth]{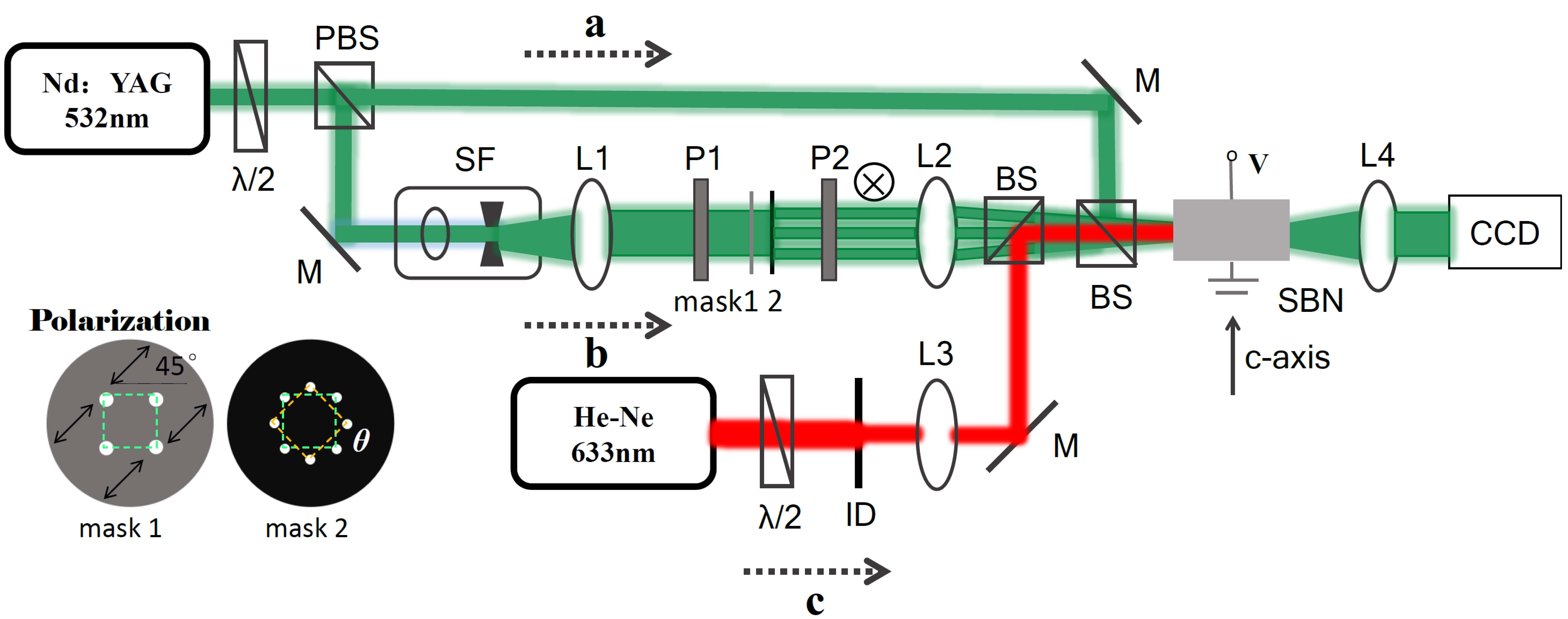}
 \caption{Experimental setup. $ \lambda/2 $, half-wave plate; PBS, polarizing beam splitter; SF, spatial filter; L, lens; BS, beam splitter; ID, iris diaphragm; M, mirror; P,   Polarizer; SBN, strontium barium niobate crystal; CCD, charged-coupled device. Mask 2 is an amplitude mask to produce two group of sub-lattices with a rotation angle $ \theta $, and Mask 1 is made of a polarizer film.}
 \label{fig:experiment}
 \end{figure}

\begin{figure}[ht]
\includegraphics[width=0.9\textwidth]{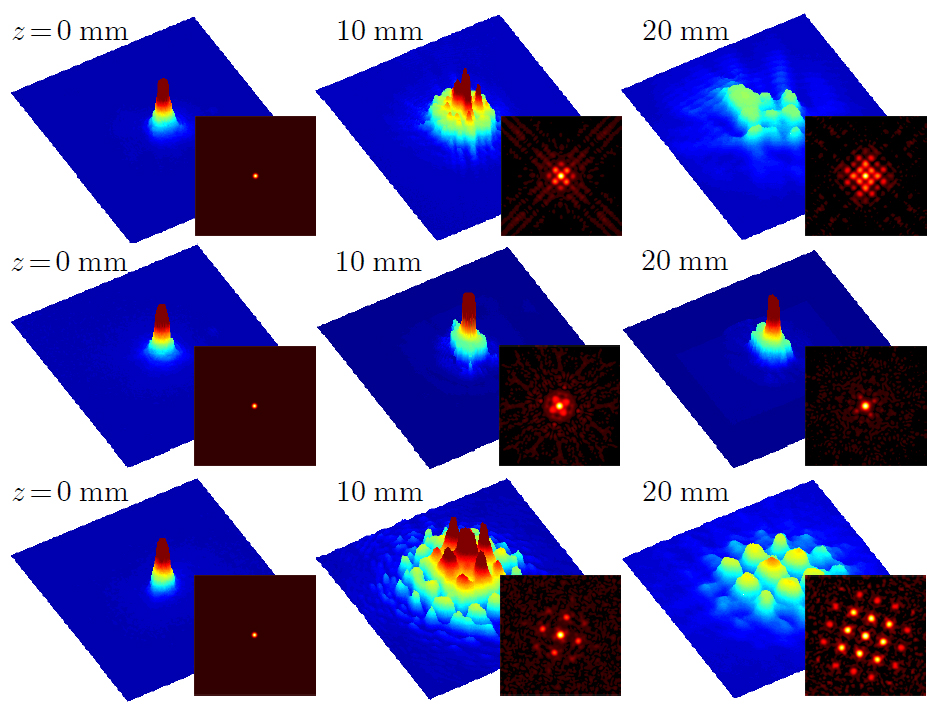}
\caption{Experimentally observed intensity distributions of the probe beam (color-surface plots) and corresponding theoretically calculated distributions (insets), at different propagation distances $z$, for $ \tan\theta=3^{-1/2}$, $ p_{2} =0.1$, which falls below the LDT point (top row), $\tan\theta=3^{-1/2}$, $p_{2} =1 $, which falls above LDT point (middle row), and $ \tan\theta=3/4 $, $ p_{2} =1$ (bottom row). The first two rows correspond to the incommensurable Pythagorean lattice shown in the central column of Fig.~1 of the main text. The third row corresponds to the commensurable lattice shown in the last column of Fig.~1 of the main text.}
 \label{fig:propexperiment}
\end{figure}
 
 \bigskip
 
\begin{figure}[ht]
\centering
\includegraphics[width=0.9\textwidth]{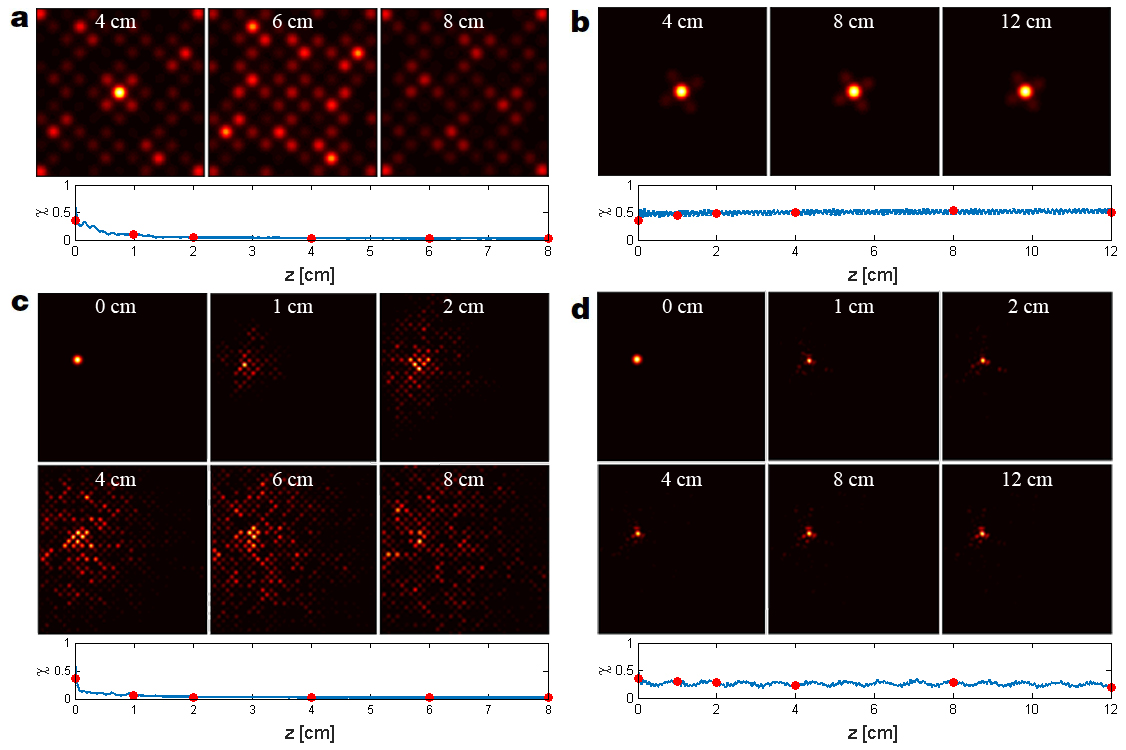}
\caption{(a),(b) Numerical simulations of the light beam propagation in the incommensurable moir\'e lattice for central excitation, corresponding to top and middle rows of Extended Data Fig. 2, but for larger distances, notably exceeding sample length. (c),(d) Similar numerical results, but for off-center excitation position in moir\'e lattice. Parameter $p_2=0.1$ for (a),(c) and $p_2=1.0$ for (b),(d), while rotation angle $\theta=\pi/6$. In all cases Gaussian beam exciting a single site of the potential is assumed.}
 \label{fig:propsimulation}
 \end{figure}

\renewcommand\tablename{Extended Data Tab.}
\begin{table}[ht]
\begin{tabular}{l|l|l|l}
Moir\'e lattice $I({\bm r})$& Sublattice $V({\bm r})$ & Diophantine equation  & $\tan\theta$
\\
\hline
Pythagorean & $\cos(2x)+\cos(2y)$ & $a^2+b^2=c^2$ & $b/a$
\\ 
hexagonal &  $\sum_{n=1}^3\cos\left[2 (x\cos\theta_n+y\sin\theta_n)\right]$ &  $a^2+b^2+ab=c^2$ & $\sqrt{3}b/(2a+b)$
\end{tabular}
\caption{Characteristics of the moir\'e lattices used in experiments. For hexagonal lattices $\theta_{1}=0$, $\theta_2=2\pi/3$, and $\theta_3=4\pi/3$.}
\label{tab:one}
\end{table}




\end{document}